\documentclass[aps,twocolumn,showpacs,superscriptaddress,preprintnumbers]{revtex4-2}
\usepackage[utf8]{inputenc}
\usepackage{amsmath,amsfonts,amssymb}
\usepackage{graphicx}
\usepackage{float}
\usepackage{subfig}
\usepackage{xcolor}
\usepackage{xspace}

\newcommand{\unit}[1]{\ensuremath{\mathrm{\,#1}}\xspace}
\newcommand{\e}{\unit{e^{-}}}

\newcommand{\fer}[1]{\textcolor{black}{#1}}

\begin{document}
%\title{Smart readout of nondestructive image sensors with single photon-electron sensitivity}
\title{Smart readout of nondestructive image sensors with \fer{single-photon sensitivity}}

\author{Fernando Chierchie}
\affiliation{Instituto de Inv. en Ing. Eléctrica ``Alfredo Desages'' (IIIE), Dpto. de Ing. Eléctrica y de Computadoras. CONICET and Universidad Nacional del Sur (UNS), Bahía Blanca, Argentina}
\author{Guillermo Fernandez Moroni}
\email[Corresponding author: ]{gfmoroni@fnal.gov}
\affiliation{Fermi National Accelerator Laboratory, Batavia IL, United States}
\author{Leandro Stefanazzi}
\affiliation{Fermi National Accelerator Laboratory, Batavia IL, United States}
\author{Eduardo Paolini}
\affiliation{Instituto de Inv. en Ing. Eléctrica ``Alfredo Desages'' (IIIE), Dpto. de Ing. Eléctrica y de Computadoras. CONICET and Universidad Nacional del Sur (UNS), Bahía Blanca, Argentina}
\author{Javier Tiffenberg}
\affiliation{Fermi National Accelerator Laboratory, Batavia IL, United States}
\author{Juan Estrada}
\affiliation{Fermi National Accelerator Laboratory, Batavia IL, United States}
\author{Gustavo Cancelo}
\affiliation{Fermi National Accelerator Laboratory, Batavia IL, United States}
\author{Sho Uemura}
\affiliation{School of Physics and Astronomy, Tel-Aviv University, Tel-Aviv, Israel}

%\begin{abstract}
%In this paper we present a smart readout technique for single-photon-electron sensitivity image sensors with nondestructive charge readout to allow its use in visible light and other applications that require faster readout times. 
%The method optimizes the readout noise and time either statically, by defining an arbitrary number of regions of interest (ROI) in the array, or dynamically, depending on the charge or energy of interest (EOI) in the pixel. This technique is tested in a Skipper CCD showing that it is possible to obtain deep sub-electron noise, and therefore, high resolution of quantized charge, while dynamically changing the readout noise of the sensor. These faster, low noise readout techniques show that  the skipper CCD is a competitive technology even where other technologies such as Electron Multiplier Charge Coupled Devices (EMCCD), silicon  photo multipliers, etc. are currently used. We discuss several scientific applications that could benefit from the proposed technique.
%\end{abstract}

\begin{abstract}
\fer{Image sensors with nondestructive charge readout provide single-photon or single-electron sensitivity, but at the cost of long readout times. We present a smart readout technique to allow the use of these sensors in visible-light and other applications that require faster readout times.}
The method optimizes the readout noise and time \fer{by changing the number of times pixels are read out} either statically, by defining an arbitrary number of regions of interest (ROI) in the array, or dynamically, depending on the charge or energy of interest (EOI) in the pixel. This technique is tested in a Skipper CCD showing that it is possible to obtain deep sub-electron noise, and therefore, high resolution of quantized charge, while dynamically changing the readout noise of the sensor. These faster, low noise readout techniques show that the skipper CCD is a competitive technology even where other technologies such as Electron Multiplier Charge Coupled Devices (EMCCD), silicon  photo multipliers, etc. are currently used. \fer{This technique could allow skipper CCDs to benefit new astronomical instruments, quantum imaging, exoplanet search and study, and quantum metrology.}
\end{abstract}

\maketitle
\section{Introduction}
%\noindent\textbf{INTRODUCTION}
Single-photon and -electron resolution semiconductor sensors have proven to be a major scientific breakthrough overcoming the limitation imposed by readout noise \cite{Simoen_1999, janesick_1990,boukhayma2017ultra}. Some technologies, such as Electron Multiplying Charge Coupled Device (EMCCD) \cite{Hynecek2003} or silicon photomultipliers \cite{BUZHAN2003}, are based on charge multiplication. More recent ones use nondestructive readout techniques to average several observations of the collected charge \cite{Tiffenberg:2017aac, stefanov2020simulations, Marshall_2019}.  Arbitrary precision is obtained at the expense of increasing the number of samples ($N$ from here) and thus the readout time. Assuming independent measurements, the noise is reduced following \cite{skipper_2012} $\sigma_0/\sqrt{N}$ where $\sigma_0$ is the the root mean squared error for one measurement of the charge. In particular, the Skipper CCD \cite{skipper_2012, Tiffenberg:2017aac} uses a floating sense node to isolate the charge packet from the first amplification stage which allows to make multiple measurements, using the correlated double sampling (CDS) method \cite{janesick2001scientific}, to get single-electron resolution pixel readout. In recent years, many applications such as dark matter searches \cite{Barak2020}, neutrino detection \cite{violeta2020}, and study of properties of semiconductor materials \cite{Rodrigues_2020} have exploited this capability, but others such as quantum imaging \cite{estrada2021ghost}, astronomical terrestrial instruments \cite{Drlica_2020},  satellite missions for exoplanet searches \cite{RauscherNASA2019}, and sub-shot-noise microscopy \cite{Samantaray2017}, remain inaccessible for the Skipper CCD due to the long readout time.

The readout noise is not always the limiting factor. Other processes produce statistical fluctuations that are added in quadrature with the electronic noise and contribute to the total uncertainty. Among these processes we can mention intrinsic factors of the sensor like quantum efficiency, leakage current, charge transfer and collection inefficiencies, crystal ionization mechanism, etc; and extrinsic factors, like the Poisson statistics of photon arrival, natural background ionizing particles, etc. \cite{Groom1999,janesick2001scientific}. When these dominate, there is no benefit to reducing the readout noise by increasing the readout time. \fer{In this article a smart readout technique to reduce readout time by changing the readout noise based on available information for the  specific application is presented.}  Experimental results using an Skipper CCD are reported, but the technique may be also applied to any existing or future sensors with nondestructive readout (either with active or passive pixels) to adapt either the readout time or the dynamic range of the measuring system. 

\section{Description of the technique}
\begin{figure}[t]
    \centering
    \includegraphics[width=0.45\textwidth]{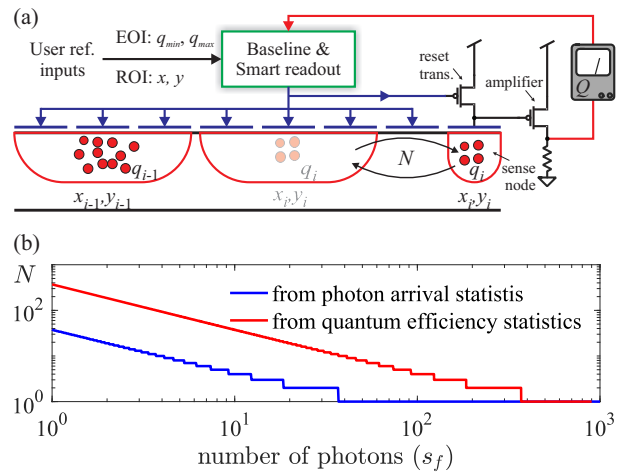}
    \caption{a) Conceptual design; b) number of samples per pixel \fer{$N$ for same fractional contribution from readout noise to photon uncertainty as a function of $s_f$; assuming $k=0.95$, $QE = 0.9$ and $\sigma_0=2$ electrons, or equivalently photons.}}
    \label{fig:nsamp_adaptation}
    \label{fig:conceptual_technique}
\end{figure}

%\noindent\textbf{CONCEPTUAL DESCRIPTION OF THE TECHNIQUE}
%
\label{sec:ConceptualTechnique}
Figure \ref{fig:conceptual_technique}(a) shows a conceptual diagram. The block ``baseline and smart readout'' confers intelligence to the output stage of the CCD to perform an adaptive modification of the number of measurements $N$ of the charge $q$ in each pixel using information of the sensor's parameters, the physics source of interest (reference inputs) and information of the current pixel. The easier strategy is to optimize $N$ according to the position of the pixel in the array ($x,y$). If the incoming photon flux illuminates a specific region, $N$ can be increased for \fer{that region}, while faster readout (and higher noise) can be used for the remaining area. This strategy adjusts the readout time and noise (or equivalently, adjusts the dynamic range) based on \textit{regions of interest} (ROI) which are previously known for the application. Another approach is to update $N$ depending on the range of charge ($q_{min},q_{max}$) or deposited energy being measured, i.e. based on the \textit{energy of interest} (EOI). 

For example, typical applications currently limited for the Skipper CCD involve visible or infrared light detection. These systems are intrinsically limited by photon statistics, which can be described by a binomial distribution (attributed to the collection efficiency of the detector) together with a Poisson distribution (attributed to photon arrival statistics).

Two scenarios are considered to show the EOI strategy: 1) a system limited by Poisson uncertainty of photon arrival, as any astronomical instrument \cite{janesick2007photon}, assuming ideal collection of photons in the sensor; 2) a system limited by photon detection uncertainty, due to quantum efficiency ($QE<1$), following a binomial distribution and assuming no Poisson arrival uncertainty, as expected in applications using entangled photons \cite{brida_2010}. In both systems, the signal to noise ratio (SNR) is $s_{f}/\sqrt{ \sigma_f^2 + \sigma_0^2/N}$, where $s_{f}$ is the expected number of collected photons and $\sigma_f$ is \fer{the standard deviation in the expected number of collected photons} ($\sigma_f=\sqrt{s_f}$ and $\sigma_f=\sqrt{s_f(1-QE)}$ for each example respectively). The number of samples per pixel $N$ can be tuned to get a desired SNR for each number of collected photons (or equivalently collected charge). \fer{If $N$ is adjusted to produce a SNR for each pixel equal to $k$ times ($k<1$) its SNR without readout noise ($\sigma_0$=0), so that the fractional contribution of readout noise to photon uncertainty is the same for every pixel, then,}
\begin{equation}
     N = \frac{\sigma_0^2k^2}{\sigma_f^2(1- k^2)}, 
     \hspace{2mm}
 \begin{cases}
  \sigma_f^2= s_{f}     &\text{Poisson} \\
  \sigma_f^2= s_f(1-QE) &\text{Binomial} 
 \end{cases}
\label{eq:SNR}
\end{equation}
%
%Figure \ref{fig:nsamp_adaptation}(b) shows $N$ as a function of $s_f$ for both examples assuming $k=0.95$ (other values for $k$ could be chosen depending on the application), $QE = 0.9$ and a readout noise of $\sigma_0=2$ electrons, or equivalently photons. 

\fer{As shown in Fig. \ref{fig:nsamp_adaptation}(b), increasing $N$ significantly decreases the uncertainty on the number of collected photons for pixels collecting fewer than $40$ photons for Poisson statistics and $400$ for binomial statistics.}
\fer{For pixels with more than this number of photons, (\ref{eq:SNR}) finds that $N=1$ is sufficient to meet the goal defined by $k$.}

This shows that an smart strategy has a big potential in reducing the readout time and tuning the dynamic range of the system for pixels with relatively small charge packets ($0.1$\% and $0.4$\% of the full pixel capacity, as tested for similar devices \cite{Diehl_2008}) which demands larger $N$ values to meet the SNR requirement. 
\fer{For a general application, assuming a uniform distribution of pixel charge values between 1\e and $100\times10^3$\e, the result in Fig. \ref{fig:nsamp_adaptation}(b) gives that the readout time with a smart strategy is $2.6\%$ and $0.26\%$ that of the non-smart strategy (all pixels read out with the highest $N$ value), in the Poisson and binomial scenarios respectively.}

\section{Implementation challenges}
%\noindent\textbf{IMPLEMENTATION CHALLENGES AND SOLUTIONS}
\label{Sec:Baseline}

Although changing the number of samples $N$ read per pixel seems straightforward, this requires changing the clock signals (twenty in our case) of the CCD ``on the fly.'' Because the CCD is a highly coupled device and the voltage swings of the clocks are on the order of tens of volts, changes to the clocks cause variations in the baseline of the video signal that, if not treated properly, introduce a systematic error in the determination of the pixel charge. Since the sensitivity of the CCD is in the order of 2$\mu$V$/e^-$, this imposes a sub-ppm control of the errors. We develop a calibration technique that can be performed on- and off-line so that $N$ can be changed without increasing the systematic error due to baseline variation.

\begin{figure}
    \centering
    \includegraphics[width=0.45\textwidth]{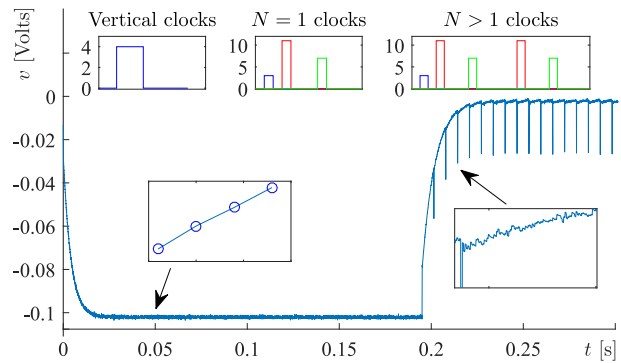} 
    \caption{Baseline of raw video signal in volts. \fer{Part of one row, the first  $3039$ pixels use $N=1$ and the others $N=100$. At the top, insets illustrate clock sequences and their voltage swings. At the bottom, zoomed regions of the baseline corresponding to the samples of a single pixel read with $N=1$ at $t\approx 0.05$ and with $N=100$ at $t\approx 0.215$.}}
    \label{fig:baselineVolts}
\end{figure}

Figure \ref{fig:baselineVolts} shows a measurement of the baseline changes in the raw video signal, before pixel computation, due to the changing clocks signals applied to a skipper CCD. The exponential decay at $t=0$ is caused by the vertical clocks applied before $t=0$. The plateau that follows is caused by many consecutive pixels read out with $N=1$. The slope within a single pixel, seen in the inset, will cause a non-zero pixel value, an effect that is present in every CCD.
The exponential at $t\approx0.2$ is caused by a change in the readout mode, from $N=1$ to $N=100$. The downward glitches seen for $t>0.2$ correspond to the first measurement of each pixel, which has a different clocking compared to the next $99$ measurements. Figure \ref{fig:baselineVolts} shows \fer{a change of $0.1$ V in the output signal base level (baseline), which exceeds by $5\times10^4$ times the expected signal for one electron ($\approx$ 2$\mu$V).}

Baseline is present in any CCD; it is usually estimated by taking an overscan region \fer{(empty pixels obtained by reading more charge packets from the serial register than there are active pixels)}, and subtracted from each image \cite{Roundy1997Baseline}. A similar approach can be used for the ROI strategy \cite{chierchie2020smartreadout}, though this requires extra calibration time to acquire empty images for each ROI. On the other hand, the EOI strategy changes $N$ online and requires corrected pixel values to be available during readout. Therefore it is mandatory to have a baseline compensation technique that can be applied online.
%This compensation strategy can no longer be used when $N$ is changed online (EOI strategy). Therefore it is mandatory to have a baseline compensation technique that does not rely on taking several empty images.

%
We developed a baseline correction technique based on the superposition of the effects of a group of control signals on the output video signal. An identification procedure is performed only once, and the baseline correction can be computed (online and offline) as the superposition of the calibrated effects for any readout sequence, either under ROI or EOI. The identification procedure is performed in the pixel values and not in the raw video signal, thus simplifying the implementation as discussed in the supplemental material \cite{SupplementalMat}.

\section{Experimental results}
%\noindent\textbf{EXPERIMENTAL RESULTS}
\label{sec:experimental}
The experimental proof of concept of the technique is done switching between $N=1$ and $N=500$ for ROI and EOI experiments. $N=500$ results in deep sub-electron noise operation and therefore any artifact introduced by the proposed adaptive readout would impact the measured total noise. Also, the jump between $N=1$ or $N=500$ produces a large change in the baseline allowing to test the capability of the readout routine to compensate for these perturbations.

Inside a dewar, the skipper-CCD is operated at high vacuum (approximately $ 10^{-4}$ mbar) and at a temperature of $140$K. The sensor is a fully-depleted CCD developed by Lawrence Berkeley National Laboratory, LBNL \cite{Tiffenberg:2017aac}. The low threshold acquisition controller (LTA) \cite{cancelo2021low} is used for readout and control.

We report three experiences: 1)  ROI specified before the readout, 2) EOI experiment choosing different charge \fer{ranges} and 3) a combination of both: \fer{once a pixel is detected in EOI, a ROI to the right of that pixel is readout with large $N$ to achieve sub-electron noise.}

For the online implementation of the EOI: the first measurement of the pixel is corrected by the baseline algorithm. If the value is within the \fer{charge range} set by the user, $N-1$ further samples are taken of the same pixel. However, if the first value is outside the range, the readout sequence continues with the next pixel. The complete baseline compensation is applied to the final image. For the ROI strategy the compensation is only applied to the final image. Further details can be found in the supplemental material \cite{SupplementalMat}.

\subsubsection{Experiment with ROI}
%\noindent\textbf{Experiment with ROI}
%
\begin{figure}
    \centering
    \includegraphics[width=0.48\textwidth]{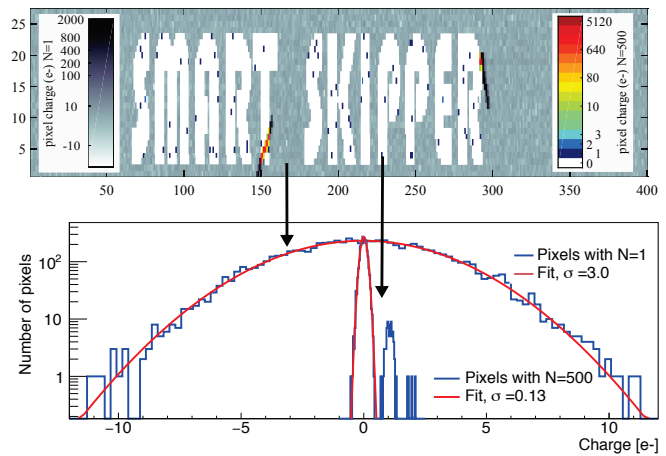}
    \caption{Measurement using ROI technique. \fer{Pixels in the words have $N= 500$ (right scale); pixels outside the words have $N=1$ (left scale).} \fer{$s_f$ was zero in most pixels, with some pixels having $s_f = 1, 2, 3$ or very large values for the two muon tracks that are observed.}}
    \label{Fig:SSWords}
\end{figure}
Fig.~\ref{Fig:SSWords} shows an image acquired with the proposed ROI technique. To show the versatility, ROIs were defined by the words ``SMART SKIPPER'': noise is 0.13\e inside the letters and 3\e outside.  Both noise measurements are obtained by Gaussian fits of the histograms, as shown in the figure. This is the theoretical expected reduction of noise when going from $N=1$ to $N=500$: $\sigma_{P_{i,skp}}=\sigma_{0}/\sqrt{N}=3/\sqrt{500} \approx 0.13$ which proves that that the baseline compensation technique does not harm the sensor charge resolution.

\subsubsection{Experiment with EOI}
%\noindent\textbf{Experiment with EOI}
%
%First we setup an experiment to obtain high charge resolution (sub-electron noise) for events with charge from zero to a few dozens of electrons.
%
\begin{figure}[t]
    \centering
     \includegraphics[width=0.48\textwidth]{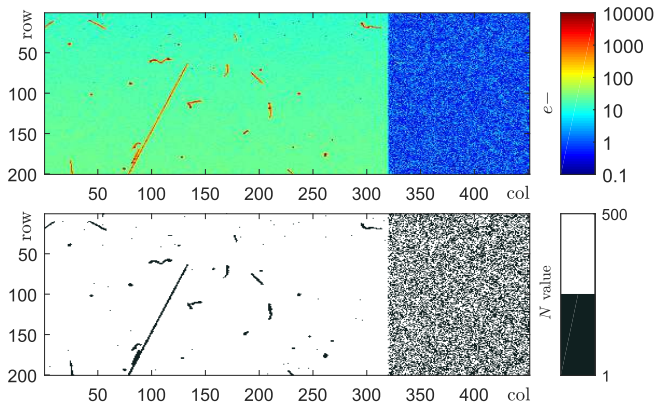}
      \caption{(Top) Image using EOI technique. (Bottom) $N$ for each pixel.}
    \label{Fig:ImageHistoandNSamp}
\end{figure}

Figure \ref{Fig:ImageHistoandNSamp} show an image taken with the proposed technique after a long exposure time to collect charge from intrinsic ionization in the sensor,
\fer{with $N=500$ for pixels with 0\e to 42\e and $N=1$ for pixels with $<0$\e or $>42$\e}. Two \fer{notable} regions are observed: part of the active region of the sensor with interacting particles and an overscan region starting in column $321$. The resulting pattern of Skipper samples $N$ is depicted at the bottom. Due to the long \fer{exposure}, most of the active region is \fer{modestly} charged and therefore read with $N=500$. The exception are the energetic muon and electron tracks, where most pixels have charge greater than 42\e and are automatically read with $N=1$. In the overscan region \fer{mostly empty pixels are present resulting in} a random pattern of $N=1$ and $N=500$ and showing the versatility of the system to change the value of $N$ in every pixel.

\begin{figure}[t]
    \centering
    \includegraphics[width=0.45\textwidth]{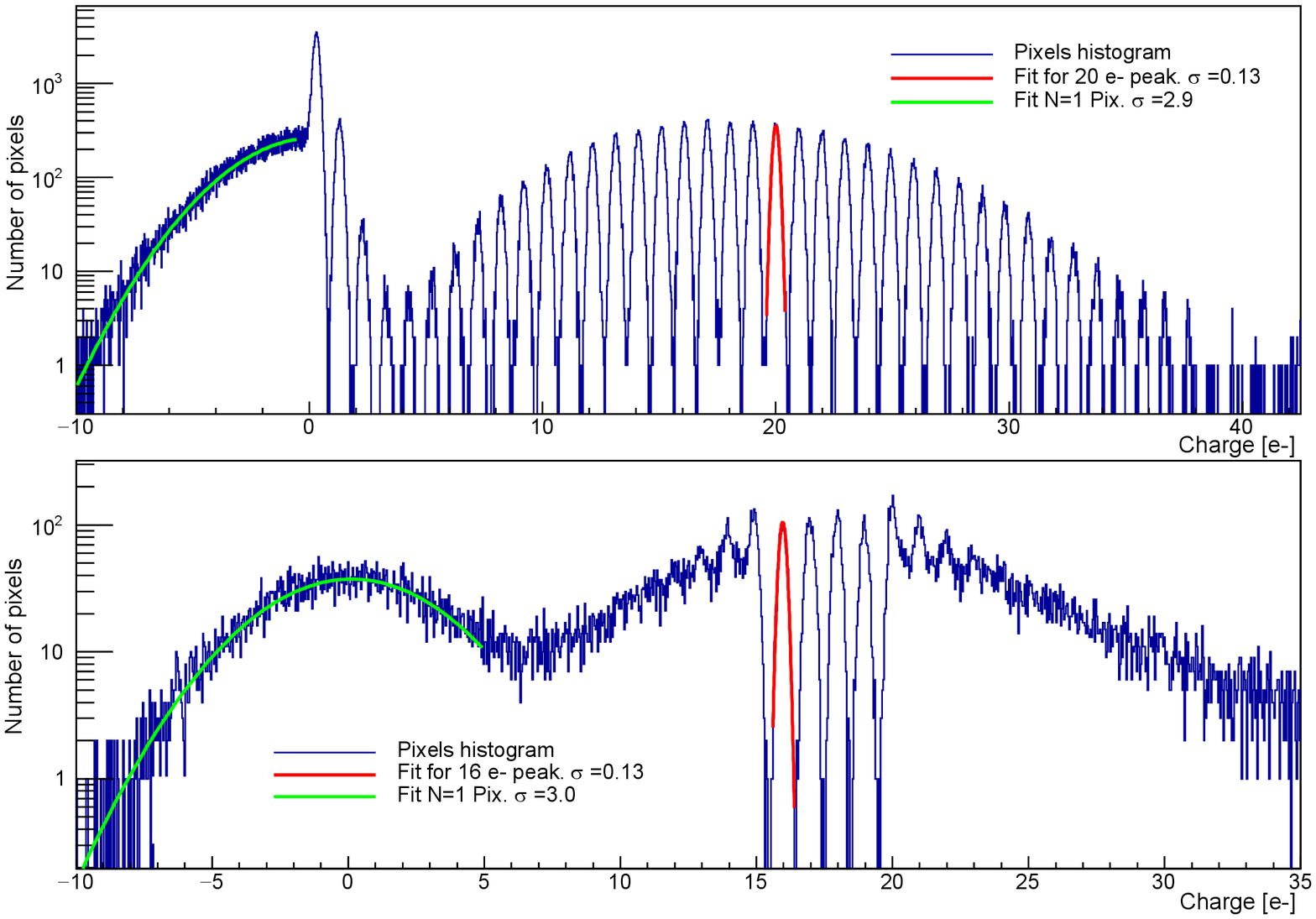} 
    \caption{Pixel histograms for EOI technique. \fer{$N= 500$ for charge ranges 0\e to 42\e (top) and 15\e to 19\e (bottom).} }
    \label{Fig:histogramsEOI}
\end{figure}

Figure \ref{Fig:histogramsEOI} shows two histograms in logarithmic scale as a result of applying the EOI technique in two different experiments. Peaks at integer numbers of electrons are clearly observed in two different charge intervals where charge quantization is achieved using $N=500$.

In both measurements, the envelope of the histograms has two distinctive bumps, one around 0\e from mostly empty pixels (mainly from the overscan) and another one centered at approximately 20\e (from the active region). The latter shows, in logarithmic scale, the characteristic Poisson distribution.

A half Gaussian distribution is observed at the left of 0\e in the histogram at the top. The green line shows a fit with  a standard deviation  $\sigma_{0}=3$\e, which is the expected readout noise for $N=1$. For $N=500$ the results are depicted in red with a fit of the 20\e peak. The standard deviation of the fit is $\sigma_{P_{i,skp}}=0.13$\e, again verifying the theoretical prediction for independent averaged measurements despite changing $N$ dynamically based on the pixel charge. The histogram at the bottom, for charge in the interval 15\e to 19\e, also shows the Gaussian fitting and the same noise performance.

\subsubsection{Experiment combining ROI and EOI}
%\noindent\textbf{Experiment combining ROI and EOI}
%
\begin{figure}[t]
    \centering
    \includegraphics[width=0.47\textwidth]{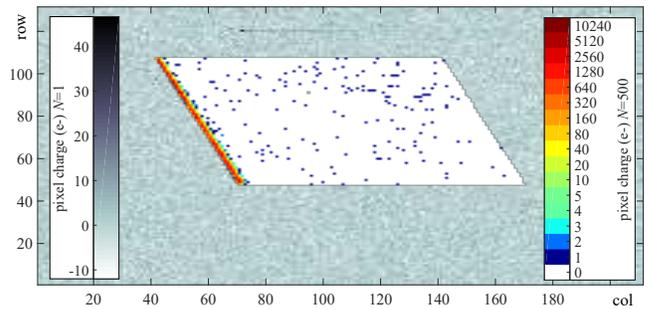} 
    \caption{Experiment combining ROI and EOI. \fer{Two color scales: on the left pixels with $N=1$ and a noise of $3$\e, and on the right pixels with $N=500$ and a readout noise of $0.13$\e (and therefore quantized charge).}}
    \label{Fig:Muon}
\end{figure}
We combine both ROI and EOI techniques using the ionization produced by a muon track to trigger sub-electron pixel measurement.

The charge \fer{range is} set between $52$\e and $6250$\e to avoid false trigger from dark current generation. If the charge of a pixel \fer{is in the range}, the current and the following $99$ pixels are read $N=500$ times, independent of their charge value.

Figure \ref{Fig:Muon} shows a fraction of the image where a muon was detected (straight line). The muon is seen mostly in red colors indicating hundreds or thousands of electrons deposited in those pixels. To the right of the muon, $99$ pixels were also read with $N=500$ samples per pixel; those pixels are observed as a white parallelogram ROI composed mostly of empty pixels (white pixels) and some pixels with $1,2,3,\cdots$\e. 

This experiment shows that it is possible to combine both techniques, which could be useful to study the charge generation around certain events of interest.

\section{Scientific Applications}
%\noindent\textbf{APPLICATIONS THAT COULD BENEFIT FROM THE PROPOSED TECHNIQUE:}
\label{sec:applications}

\subsubsection*{New astronomical instruments}
%\noindent\textbf{New astronomical instruments}
 %
The potential of the sub-electron noise of the Skipper CCD is being explored for terrestrial astronomy \cite{Tiffenberg:2017aac,Drlica_2020} for signals with small SNR. Acquisitions with limited exposure time naturally produce low SNR observations where the impact of readout noise is high. Authors claim that a reduction of $100$ in the readout time is still required, which is similar to the time reduction obtained by EOI in previous section in systems limited by photon arrival statistics. Moreover, for spectrography instruments where spectra are well defined areas in the sensor \cite{howell2006handbook}, a ROI strategy could further improve the readout speed. \fer{To quantitatively address this scenario we obtained a real image from the LDSS-3 [T. Diehl 2018, private communication], a high efficiency optical wide-field imager and multi-slit spectrograph \cite{LDSS3}. The value $N$ for each pixel based on the Poisson uncertainty was computed and resulted in a reduction of the readout time of a factor of $100$.}

\subsubsection*{Quantum imaging techniques}
%\noindent\textbf{Quantum imaging techniques}
%
%Light revels quantum properties at room temperature and this has motivated it use as an experimental test bench for quantum mechanic theory and developing of novel applications.  
A pair of photons can exhibit spatial, spectral and polarization quantum entanglement. Spatial entanglement has been extensively explored for quantum communication \cite{YUAN20101}. Spontaneous parametric down-conversion crystals have eased the production of entangled pairs over a large number of positions \cite{malygin_1985} \fer{for example for ghost imaging \cite{Padgett_2017,Pittman_1995}. Intensity correlation can be used for several other imaging techniques \cite{defienne_2010} such as fluorescence correlation spectroscopy \cite{Bestvater_2010}. }Two-dimensional semiconductor devices provide a good sensor solution for these applications \cite{Orieux_2017}. In particular, the single photon counting capability and large quantum efficiency of the Skipper CCD make it a promising technology in the field \cite{estrada2021ghost}. Moreover, as detailed in \cite{defienne_2010} on-chip noise sources impact the final measurable correlation of entangled photons, and therefore the EOI and ROI strategies can be used due to the high spatial and intensity correlation between entangled photons in two known regions of the image. 
% systems and therefore an EOI strategy could further help to improve the results.

\subsubsection*{Exoplanet search and study}
%\noindent\textbf{Exoplanet search and study}
%
Direct imaging space telescopes for exoplanet detection and characterization are being planned \cite{crill2018exoplanet}. One of the main goals for future missions is to search for near-infrared photons at $950$ nm from water vapor in the atmosphere of potentially habitable planets. A sensor with large quantum efficiency and sub-electron readout noise is required. The Skipper CCD has been identified by NASA as a promising technology that meets both requirements \cite{RauscherNASA2019,Rauscher_internal_2019,crill2018exoplanet}.
Moreover, the radiation hardness compared to EMCCD makes it more suitable for space missions. The main identified challenge is the slow readout time ($\approx20$ minutes for $1$ Mpixel array read by one amplifier at deep sub-electron noise). \fer{According to \cite{Stark_2019} less than 200 hundred pixels per spectral element will be needed for the mission's spectrograph. To overcome the readout time limitation one possibility is to use the ROI strategy to directly focus on the key wavelengths and meet the 20 second readout time constraint \cite{Rauscher_internal_2019}}.

%Authors in \cite{} reported that the expected signature from earth like planets is smaller than two hundred pixels. The rest of the array could be read using less energy resolution with a substantial reduction in the total readout time.

%\noindent\textbf{Electron pump for quantum metrology}
\subsubsection*{Electron pump for quantum metrology}
Recently there has been a redefinition of  the ampere by means of the charge of the electron \cite{BIMP2019,scherer2019single}. One of the technological candidates for metrology is the single-electron transistor (SET) \cite{giblin2012towards}. These quantum devices are operated at milli-Kelvin temperatures, which complicates the scaling required to achieve reasonable practical currents (in the order of $1\mu$A).
The Skipper CCD is a promising technology for the development of a current source with single-electron manipulation. Although scaling is still a challenge, one advantages is its higher temperature of operation (in the order of $100$ Kelvin). To reach an stable average current a smart readout technique is mandatory, since charge should be measured and drained out of the device at a rate that depends on the actual charge packet measurement \cite{CCDFuente}.

\begin{acknowledgments}
	We thank the SiDet team at Fermilab for the support on the operations of CCDs and Skipper-CCDs, especially Kevin Kuk and Andrew Lathrop. Lawrence Berkeley National Laboratory is the developer of the fully-depleted CCD and the designer of the Skipper readout. The CCD development work was supported in part by the Director, Office of Science, of the U.S. Department of Energy under No. DE-AC02-05CH11231.
\end{acknowledgments}

%\section{Conclusions}
%\noindent\textbf{Conclusions}
%
%We have presented and experimentally demonstrated a novel technique to readout image sensors with nondestructive readout stages. The proposed approach achieves single photon-electron sensitivity allowing to adjust the noise in each pixel based on the physical quantity being measured, either depending on the pixel region or pixel deposited charge or energy. The proposed smart readout allows the use of the Skipper-CCD, in applications that have a tighter time budget. The Skipper CCD has the potential of overcoming other single electron-photon technologies due to its high energy resolution over the entire dynamic range.

% References
%\bibliography{main.bib} 

\begin{thebibliography}{41}%
	\makeatletter
	\providecommand \@ifxundefined [1]{%
		\@ifx{#1\undefined}
	}%
	\providecommand \@ifnum [1]{%
		\ifnum #1\expandafter \@firstoftwo
		\else \expandafter \@secondoftwo
		\fi
	}%
	\providecommand \@ifx [1]{%
		\ifx #1\expandafter \@firstoftwo
		\else \expandafter \@secondoftwo
		\fi
	}%
	\providecommand \natexlab [1]{#1}%
	\providecommand \enquote  [1]{``#1''}%
	\providecommand \bibnamefont  [1]{#1}%
	\providecommand \bibfnamefont [1]{#1}%
	\providecommand \citenamefont [1]{#1}%
	\providecommand \href@noop [0]{\@secondoftwo}%
	\providecommand \href [0]{\begingroup \@sanitize@url \@href}%
	\providecommand \@href[1]{\@@startlink{#1}\@@href}%
	\providecommand \@@href[1]{\endgroup#1\@@endlink}%
	\providecommand \@sanitize@url [0]{\catcode `\\12\catcode `\$12\catcode
		`\&12\catcode `\#12\catcode `\^12\catcode `\_12\catcode `\%12\relax}%
	\providecommand \@@startlink[1]{}%
	\providecommand \@@endlink[0]{}%
	\providecommand \url  [0]{\begingroup\@sanitize@url \@url }%
	\providecommand \@url [1]{\endgroup\@href {#1}{\urlprefix }}%
	\providecommand \urlprefix  [0]{URL }%
	\providecommand \Eprint [0]{\href }%
	\providecommand \doibase [0]{http://dx.doi.org/}%
	\providecommand \selectlanguage [0]{\@gobble}%
	\providecommand \bibinfo  [0]{\@secondoftwo}%
	\providecommand \bibfield  [0]{\@secondoftwo}%
	\providecommand \translation [1]{[#1]}%
	\providecommand \BibitemOpen [0]{}%
	\providecommand \bibitemStop [0]{}%
	\providecommand \bibitemNoStop [0]{.\EOS\space}%
	\providecommand \EOS [0]{\spacefactor3000\relax}%
	\providecommand \BibitemShut  [1]{\csname bibitem#1\endcsname}%
	\let\auto@bib@innerbib\@empty
	%</preamble>
	\bibitem [{\citenamefont {Simoen}\ and\ \citenamefont
		{Claeys}(1999)}]{Simoen_1999}%
	\BibitemOpen
	\bibfield  {author} {\bibinfo {author} {\bibfnamefont {E.}~\bibnamefont
			{Simoen}}\ and\ \bibinfo {author} {\bibfnamefont {C.}~\bibnamefont
			{Claeys}},\ }\href {\doibase https://doi.org/10.1016/S0038-1101(98)00322-0}
	{\bibfield  {journal} {\bibinfo  {journal} {Solid-State Electronics}\
		}\textbf {\bibinfo {volume} {43}},\ \bibinfo {pages} {865} (\bibinfo {year}
		{1999})}\BibitemShut {NoStop}%
	\bibitem [{\citenamefont {Janesick}\ \emph {et~al.}(1990)\citenamefont
		{Janesick}, \citenamefont {Elliott}, \citenamefont {Dingiziam}, \citenamefont
		{Bredthauer}, \citenamefont {Chandler}, \citenamefont {Westphal},\ and\
		\citenamefont {Gunn}}]{janesick_1990}%
	\BibitemOpen
	\bibfield  {author} {\bibinfo {author} {\bibfnamefont {J.~R.}\ \bibnamefont
			{Janesick}}, \bibinfo {author} {\bibfnamefont {T.~S.}\ \bibnamefont
			{Elliott}}, \bibinfo {author} {\bibfnamefont {A.}~\bibnamefont {Dingiziam}},
		\bibinfo {author} {\bibfnamefont {R.~A.}\ \bibnamefont {Bredthauer}},
		\bibinfo {author} {\bibfnamefont {C.~E.}\ \bibnamefont {Chandler}}, \bibinfo
		{author} {\bibfnamefont {J.~A.}\ \bibnamefont {Westphal}}, \ and\ \bibinfo
		{author} {\bibfnamefont {J.~E.}\ \bibnamefont {Gunn}},\ }in\ \href {\doibase
		10.1117/12.19452} {\emph {\bibinfo {booktitle} {Charge-Coupled Devices and
				Solid State Optical Sensors}}},\ Vol.\ \bibinfo {volume} {1242},\ \bibinfo
	{editor} {edited by\ \bibinfo {editor} {\bibfnamefont {M.~M.}\ \bibnamefont
			{Blouke}}},\ \bibinfo {organization} {International Society for Optics and
		Photonics}\ (\bibinfo  {publisher} {SPIE},\ \bibinfo {year} {1990})\ pp.\
	\bibinfo {pages} {223 -- 237}\BibitemShut {NoStop}%
	\bibitem [{\citenamefont {Boukhayma}(2017)}]{boukhayma2017ultra}%
	\BibitemOpen
	\bibfield  {author} {\bibinfo {author} {\bibfnamefont {A.}~\bibnamefont
			{Boukhayma}},\ }\href {https://books.google.com.ar/books?id=ippADwAAQBAJ}
	{\emph {\bibinfo {title} {Ultra Low Noise CMOS Image Sensors}}},\ Springer
	Theses\ (\bibinfo  {publisher} {Springer International Publishing},\ \bibinfo
	{year} {2017})\BibitemShut {NoStop}%
	\bibitem [{\citenamefont {Hynecek}\ and\ \citenamefont
		{Nishiwaki}(2003)}]{Hynecek2003}%
	\BibitemOpen
	\bibfield  {author} {\bibinfo {author} {\bibfnamefont {J.}~\bibnamefont
			{Hynecek}}\ and\ \bibinfo {author} {\bibfnamefont {T.}~\bibnamefont
			{Nishiwaki}},\ }\href {\doibase 10.1109/TED.2002.806962} {\bibfield
		{journal} {\bibinfo  {journal} {IEEE Transactions on Electron Devices}\
		}\textbf {\bibinfo {volume} {50}},\ \bibinfo {pages} {239} (\bibinfo {year}
		{2003})}\BibitemShut {NoStop}%
	\bibitem [{\citenamefont {Buzhan}\ \emph {et~al.}(2003)\citenamefont {Buzhan},
		\citenamefont {Dolgoshein}, \citenamefont {Filatov}, \citenamefont {Ilyin},
		\citenamefont {Kantzerov}, \citenamefont {Kaplin}, \citenamefont {Karakash},
		\citenamefont {Kayumov}, \citenamefont {Klemin}, \citenamefont {Popova},\
		and\ \citenamefont {Smirnov}}]{BUZHAN2003}%
	\BibitemOpen
	\bibfield  {author} {\bibinfo {author} {\bibfnamefont {P.}~\bibnamefont
			{Buzhan}}, \bibinfo {author} {\bibfnamefont {B.}~\bibnamefont {Dolgoshein}},
		\bibinfo {author} {\bibfnamefont {L.}~\bibnamefont {Filatov}}, \bibinfo
		{author} {\bibfnamefont {A.}~\bibnamefont {Ilyin}}, \bibinfo {author}
		{\bibfnamefont {V.}~\bibnamefont {Kantzerov}}, \bibinfo {author}
		{\bibfnamefont {V.}~\bibnamefont {Kaplin}}, \bibinfo {author} {\bibfnamefont
			{A.}~\bibnamefont {Karakash}}, \bibinfo {author} {\bibfnamefont
			{F.}~\bibnamefont {Kayumov}}, \bibinfo {author} {\bibfnamefont
			{S.}~\bibnamefont {Klemin}}, \bibinfo {author} {\bibfnamefont
			{E.}~\bibnamefont {Popova}}, \ and\ \bibinfo {author} {\bibfnamefont
			{S.}~\bibnamefont {Smirnov}},\ }\href {\doibase
		https://doi.org/10.1016/S0168-9002(03)00749-6} {\bibfield  {journal}
		{\bibinfo  {journal} {Nuclear Instruments and Methods in Physics Research
				Section A: Accelerators, Spectrometers, Detectors and Associated Equipment}\
		}\textbf {\bibinfo {volume} {504}},\ \bibinfo {pages} {48} (\bibinfo {year}
		{2003})},\ \bibinfo {note} {proceedings of the 3rd International Conference
		on New Developments in Photodetection}\BibitemShut {NoStop}%
	\bibitem [{\citenamefont {Tiffenberg}\ \emph {et~al.}(2017)\citenamefont
		{Tiffenberg}, \citenamefont {Sofo-Haro}, \citenamefont {Drlica-Wagner},
		\citenamefont {Essig}, \citenamefont {Guardincerri}, \citenamefont {Holland},
		\citenamefont {Volansky},\ and\ \citenamefont {Yu}}]{Tiffenberg:2017aac}%
	\BibitemOpen
	\bibfield  {author} {\bibinfo {author} {\bibfnamefont {J.}~\bibnamefont
			{Tiffenberg}}, \bibinfo {author} {\bibfnamefont {M.}~\bibnamefont
			{Sofo-Haro}}, \bibinfo {author} {\bibfnamefont {A.}~\bibnamefont
			{Drlica-Wagner}}, \bibinfo {author} {\bibfnamefont {R.}~\bibnamefont
			{Essig}}, \bibinfo {author} {\bibfnamefont {Y.}~\bibnamefont {Guardincerri}},
		\bibinfo {author} {\bibfnamefont {S.}~\bibnamefont {Holland}}, \bibinfo
		{author} {\bibfnamefont {T.}~\bibnamefont {Volansky}}, \ and\ \bibinfo
		{author} {\bibfnamefont {T.-T.}\ \bibnamefont {Yu}},\ }\href {\doibase
		10.1103/PhysRevLett.119.131802} {\bibfield  {journal} {\bibinfo  {journal}
			{Phys. Rev. Lett.}\ }\textbf {\bibinfo {volume} {119}},\ \bibinfo {pages}
		{131802} (\bibinfo {year} {2017})},\ \Eprint
	{http://arxiv.org/abs/1706.00028} {1706.00028} \BibitemShut {NoStop}%
	\bibitem [{\citenamefont {Stefanov}\ \emph {et~al.}(2020)\citenamefont
		{Stefanov}, \citenamefont {Prest}, \citenamefont {Downing}, \citenamefont
		{George}, \citenamefont {Bezawada},\ and\ \citenamefont
		{Holland}}]{stefanov2020simulations}%
	\BibitemOpen
	\bibfield  {author} {\bibinfo {author} {\bibfnamefont {K.~D.}\ \bibnamefont
			{Stefanov}}, \bibinfo {author} {\bibfnamefont {M.~J.}\ \bibnamefont {Prest}},
		\bibinfo {author} {\bibfnamefont {M.}~\bibnamefont {Downing}}, \bibinfo
		{author} {\bibfnamefont {E.}~\bibnamefont {George}}, \bibinfo {author}
		{\bibfnamefont {N.}~\bibnamefont {Bezawada}}, \ and\ \bibinfo {author}
		{\bibfnamefont {A.~D.}\ \bibnamefont {Holland}},\ }\href@noop {} {\bibfield
		{journal} {\bibinfo  {journal} {Sensors}\ }\textbf {\bibinfo {volume} {20}},\
		\bibinfo {pages} {2031} (\bibinfo {year} {2020})}\BibitemShut {NoStop}%
	\bibitem [{\citenamefont {Bautz}\ \emph {et~al.}(2019)\citenamefont {Bautz},
		\citenamefont {Burke}, \citenamefont {Cooper}, \citenamefont {Craig},
		\citenamefont {Foster}, \citenamefont {Grant}, \citenamefont {LaMarr},
		\citenamefont {Leitz}, \citenamefont {Malonis}, \citenamefont {Miller},
		\citenamefont {Prigozhin}, \citenamefont {Schuette}, \citenamefont
		{Suntharalingam},\ and\ \citenamefont {Thayer}}]{Marshall_2019}%
	\BibitemOpen
	\bibfield  {author} {\bibinfo {author} {\bibfnamefont {M.~W.}\ \bibnamefont
			{Bautz}}, \bibinfo {author} {\bibfnamefont {B.~E.}\ \bibnamefont {Burke}},
		\bibinfo {author} {\bibfnamefont {M.}~\bibnamefont {Cooper}}, \bibinfo
		{author} {\bibfnamefont {D.}~\bibnamefont {Craig}}, \bibinfo {author}
		{\bibfnamefont {R.~F.}\ \bibnamefont {Foster}}, \bibinfo {author}
		{\bibfnamefont {C.~E.}\ \bibnamefont {Grant}}, \bibinfo {author}
		{\bibfnamefont {B.~J.}\ \bibnamefont {LaMarr}}, \bibinfo {author}
		{\bibfnamefont {C.}~\bibnamefont {Leitz}}, \bibinfo {author} {\bibfnamefont
			{A.}~\bibnamefont {Malonis}}, \bibinfo {author} {\bibfnamefont {E.~D.}\
			\bibnamefont {Miller}}, \bibinfo {author} {\bibfnamefont {G.}~\bibnamefont
			{Prigozhin}}, \bibinfo {author} {\bibfnamefont {D.}~\bibnamefont {Schuette}},
		\bibinfo {author} {\bibfnamefont {V.}~\bibnamefont {Suntharalingam}}, \ and\
		\bibinfo {author} {\bibfnamefont {C.}~\bibnamefont {Thayer}},\ }\href
	{\doibase 10.1117/1.JATIS.5.2.021015} {\bibfield  {journal} {\bibinfo
			{journal} {Journal of Astronomical Telescopes, Instruments, and Systems}\
		}\textbf {\bibinfo {volume} {5}},\ \bibinfo {pages} {1 } (\bibinfo {year}
		{2019})}\BibitemShut {NoStop}%
	\bibitem [{\citenamefont {Fernandez~Moroni}\ \emph {et~al.}(2012)\citenamefont
		{Fernandez~Moroni}, \citenamefont {Estrada}, \citenamefont {Cancelo},
		\citenamefont {Holland}, \citenamefont {Paolini},\ and\ \citenamefont
		{Diehl}}]{skipper_2012}%
	\BibitemOpen
	\bibfield  {author} {\bibinfo {author} {\bibfnamefont {G.}~\bibnamefont
			{Fernandez~Moroni}}, \bibinfo {author} {\bibfnamefont {J.}~\bibnamefont
			{Estrada}}, \bibinfo {author} {\bibfnamefont {G.}~\bibnamefont {Cancelo}},
		\bibinfo {author} {\bibfnamefont {S.}~\bibnamefont {Holland}}, \bibinfo
		{author} {\bibfnamefont {E.}~\bibnamefont {Paolini}}, \ and\ \bibinfo
		{author} {\bibfnamefont {H.}~\bibnamefont {Diehl}},\ }\href {\doibase
		10.1007/s10686-012-9298-x} {\bibfield  {journal} {\bibinfo  {journal}
			{Experimental Astronomy}\ }\textbf {\bibinfo {volume} {34}} (\bibinfo {year}
		{2012}),\ 10.1007/s10686-012-9298-x}\BibitemShut {NoStop}%
	\bibitem [{\citenamefont {Janesick}(2001)}]{janesick2001scientific}%
	\BibitemOpen
	\bibfield  {author} {\bibinfo {author} {\bibfnamefont {J.~R.}\ \bibnamefont
			{Janesick}},\ }\href@noop {} {\emph {\bibinfo {title} {Scientific
				charge-coupled devices}}},\ Vol.~\bibinfo {volume} {83}\ (\bibinfo
	{publisher} {SPIE press},\ \bibinfo {year} {2001})\BibitemShut {NoStop}%
	\bibitem [{\citenamefont {Barak}\ \emph {et~al.}(2020)\citenamefont {Barak},
		\citenamefont {Bloch}, \citenamefont {Cababie}, \citenamefont {Cancelo},
		\citenamefont {Chaplinsky}, \citenamefont {Chierchie}, \citenamefont
		{Crisler}, \citenamefont {Drlica-Wagner}, \citenamefont {Essig},
		\citenamefont {Estrada}, \citenamefont {Etzion}, \citenamefont {Moroni},
		\citenamefont {Gift}, \citenamefont {Munagavalasa}, \citenamefont {Orly},
		\citenamefont {Rodrigues}, \citenamefont {Singal}, \citenamefont {Haro},
		\citenamefont {Stefanazzi}, \citenamefont {Tiffenberg}, \citenamefont
		{Uemura}, \citenamefont {Volansky},\ and\ \citenamefont {Yu}}]{Barak2020}%
	\BibitemOpen
	\bibfield  {author} {\bibinfo {author} {\bibfnamefont {L.}~\bibnamefont
			{Barak}}, \bibinfo {author} {\bibfnamefont {I.~M.}\ \bibnamefont {Bloch}},
		\bibinfo {author} {\bibfnamefont {M.}~\bibnamefont {Cababie}}, \bibinfo
		{author} {\bibfnamefont {G.}~\bibnamefont {Cancelo}}, \bibinfo {author}
		{\bibfnamefont {L.}~\bibnamefont {Chaplinsky}}, \bibinfo {author}
		{\bibfnamefont {F.}~\bibnamefont {Chierchie}}, \bibinfo {author}
		{\bibfnamefont {M.}~\bibnamefont {Crisler}}, \bibinfo {author} {\bibfnamefont
			{A.}~\bibnamefont {Drlica-Wagner}}, \bibinfo {author} {\bibfnamefont
			{R.}~\bibnamefont {Essig}}, \bibinfo {author} {\bibfnamefont
			{J.}~\bibnamefont {Estrada}}, \bibinfo {author} {\bibfnamefont
			{E.}~\bibnamefont {Etzion}}, \bibinfo {author} {\bibfnamefont {G.~F.}\
			\bibnamefont {Moroni}}, \bibinfo {author} {\bibfnamefont {D.}~\bibnamefont
			{Gift}}, \bibinfo {author} {\bibfnamefont {S.}~\bibnamefont {Munagavalasa}},
		\bibinfo {author} {\bibfnamefont {A.}~\bibnamefont {Orly}}, \bibinfo {author}
		{\bibfnamefont {D.}~\bibnamefont {Rodrigues}}, \bibinfo {author}
		{\bibfnamefont {A.}~\bibnamefont {Singal}}, \bibinfo {author} {\bibfnamefont
			{M.~S.}\ \bibnamefont {Haro}}, \bibinfo {author} {\bibfnamefont
			{L.}~\bibnamefont {Stefanazzi}}, \bibinfo {author} {\bibfnamefont
			{J.}~\bibnamefont {Tiffenberg}}, \bibinfo {author} {\bibfnamefont
			{S.}~\bibnamefont {Uemura}}, \bibinfo {author} {\bibfnamefont
			{T.}~\bibnamefont {Volansky}}, \ and\ \bibinfo {author} {\bibfnamefont
			{T.-T.}\ \bibnamefont {Yu}} (\bibinfo {collaboration} {SENSEI
			Collaboration}),\ }\href {\doibase 10.1103/PhysRevLett.125.171802} {\bibfield
		{journal} {\bibinfo  {journal} {Phys. Rev. Lett.}\ }\textbf {\bibinfo
			{volume} {125}},\ \bibinfo {pages} {171802} (\bibinfo {year}
		{2020})}\BibitemShut {NoStop}%
	\bibitem [{\citenamefont {D'Olivo}\ \emph {et~al.}(2020)\citenamefont
		{D'Olivo}, \citenamefont {Bonifazi}, \citenamefont {Rodrigues},\ and\
		\citenamefont {Moroni}}]{violeta2020}%
	\BibitemOpen
	\bibfield  {author} {\bibinfo {author} {\bibfnamefont {J.~C.}\ \bibnamefont
			{D'Olivo}}, \bibinfo {author} {\bibfnamefont {C.}~\bibnamefont {Bonifazi}},
		\bibinfo {author} {\bibfnamefont {D.}~\bibnamefont {Rodrigues}}, \ and\
		\bibinfo {author} {\bibfnamefont {G.~F.}\ \bibnamefont {Moroni}},\ }in\
	\href@noop {} {\emph {\bibinfo {booktitle} {XXIX International Conference in
				Neutrino Physics, poster 521}}}\ (\bibinfo {year} {2020})\BibitemShut
	{NoStop}%
	\bibitem [{\citenamefont {Rodrigues}\ \emph {et~al.}(2020)\citenamefont
		{Rodrigues}, \citenamefont {Andersson}, \citenamefont {Cababie},
		\citenamefont {Donadon}, \citenamefont {Cancelo}, \citenamefont {Estrada},
		\citenamefont {Fernandez-Moroni}, \citenamefont {Piegaia}, \citenamefont
		{Senger}, \citenamefont {Haro} \emph {et~al.}}]{Rodrigues_2020}%
	\BibitemOpen
	\bibfield  {author} {\bibinfo {author} {\bibfnamefont {D.}~\bibnamefont
			{Rodrigues}}, \bibinfo {author} {\bibfnamefont {K.}~\bibnamefont
			{Andersson}}, \bibinfo {author} {\bibfnamefont {M.}~\bibnamefont {Cababie}},
		\bibinfo {author} {\bibfnamefont {A.}~\bibnamefont {Donadon}}, \bibinfo
		{author} {\bibfnamefont {G.}~\bibnamefont {Cancelo}}, \bibinfo {author}
		{\bibfnamefont {J.}~\bibnamefont {Estrada}}, \bibinfo {author} {\bibfnamefont
			{G.}~\bibnamefont {Fernandez-Moroni}}, \bibinfo {author} {\bibfnamefont
			{R.}~\bibnamefont {Piegaia}}, \bibinfo {author} {\bibfnamefont
			{M.}~\bibnamefont {Senger}}, \bibinfo {author} {\bibfnamefont {M.~S.}\
			\bibnamefont {Haro}},  \emph {et~al.},\ }\href@noop {} {\bibfield  {journal}
		{\bibinfo  {journal} {arXiv preprint arXiv:2004.11499}\ } (\bibinfo {year}
		{2020})}\BibitemShut {NoStop}%
	\bibitem [{\citenamefont {Estrada}\ \emph {et~al.}(2021)\citenamefont
		{Estrada}, \citenamefont {Harnik}, \citenamefont {Rodrigues},\ and\
		\citenamefont {Senger}}]{estrada2021ghost}%
	\BibitemOpen
	\bibfield  {author} {\bibinfo {author} {\bibfnamefont {J.}~\bibnamefont
			{Estrada}}, \bibinfo {author} {\bibfnamefont {R.}~\bibnamefont {Harnik}},
		\bibinfo {author} {\bibfnamefont {D.}~\bibnamefont {Rodrigues}}, \ and\
		\bibinfo {author} {\bibfnamefont {M.}~\bibnamefont {Senger}},\ }\href@noop {}
	{\enquote {\bibinfo {title} {Ghost imaging of dark particles},}\ } (\bibinfo
	{year} {2021}),\ \Eprint {http://arxiv.org/abs/2012.04707} {arXiv:2012.04707
		[hep-ph]} \BibitemShut {NoStop}%
	\bibitem [{\citenamefont {Drlica-Wagner}\ \emph {et~al.}(2020)\citenamefont
		{Drlica-Wagner}, \citenamefont {Villalpando}, \citenamefont {O'Neil},
		\citenamefont {Estrada}, \citenamefont {Holland}, \citenamefont {Kurinsky},
		\citenamefont {Li}, \citenamefont {Moroni}, \citenamefont {Tiffenberg},\ and\
		\citenamefont {Uemura}}]{Drlica_2020}%
	\BibitemOpen
	\bibfield  {author} {\bibinfo {author} {\bibfnamefont {A.}~\bibnamefont
			{Drlica-Wagner}}, \bibinfo {author} {\bibfnamefont {E.~M.}\ \bibnamefont
			{Villalpando}}, \bibinfo {author} {\bibfnamefont {J.}~\bibnamefont {O'Neil}},
		\bibinfo {author} {\bibfnamefont {J.}~\bibnamefont {Estrada}}, \bibinfo
		{author} {\bibfnamefont {S.}~\bibnamefont {Holland}}, \bibinfo {author}
		{\bibfnamefont {N.}~\bibnamefont {Kurinsky}}, \bibinfo {author}
		{\bibfnamefont {T.}~\bibnamefont {Li}}, \bibinfo {author} {\bibfnamefont
			{G.~F.}\ \bibnamefont {Moroni}}, \bibinfo {author} {\bibfnamefont
			{J.}~\bibnamefont {Tiffenberg}}, \ and\ \bibinfo {author} {\bibfnamefont
			{S.}~\bibnamefont {Uemura}},\ }in\ \href {\doibase 10.1117/12.2562403} {\emph
		{\bibinfo {booktitle} {X-Ray, Optical, and Infrared Detectors for Astronomy
				IX}}},\ Vol.\ \bibinfo {volume} {11454},\ \bibinfo {editor} {edited by\
		\bibinfo {editor} {\bibfnamefont {A.~D.}\ \bibnamefont {Holland}}\ and\
		\bibinfo {editor} {\bibfnamefont {J.}~\bibnamefont {Beletic}}},\ \bibinfo
	{organization} {International Society for Optics and Photonics}\ (\bibinfo
	{publisher} {SPIE},\ \bibinfo {year} {2020})\ pp.\ \bibinfo {pages} {210 --
		223}\BibitemShut {NoStop}%
	\bibitem [{\citenamefont {Rauscher}\ \emph {et~al.}(2019)\citenamefont
		{Rauscher}, \citenamefont {Holland}, \citenamefont {Miko},\ and\
		\citenamefont {Waczynski}}]{RauscherNASA2019}%
	\BibitemOpen
	\bibfield  {author} {\bibinfo {author} {\bibfnamefont {B.~J.}\ \bibnamefont
			{Rauscher}}, \bibinfo {author} {\bibfnamefont {S.~E.}\ \bibnamefont
			{Holland}}, \bibinfo {author} {\bibfnamefont {L.~R.}\ \bibnamefont {Miko}}, \
		and\ \bibinfo {author} {\bibfnamefont {A.}~\bibnamefont {Waczynski}},\ }in\
	\href {\doibase 10.1117/12.2536509} {\emph {\bibinfo {booktitle}
			{UV/Optical/IR Space Telescopes and Instruments: Innovative Technologies and
				Concepts IX}}},\ Vol.\ \bibinfo {volume} {11115},\ \bibinfo {editor} {edited
		by\ \bibinfo {editor} {\bibfnamefont {A.~A.}\ \bibnamefont {Barto}}, \bibinfo
		{editor} {\bibfnamefont {J.~B.}\ \bibnamefont {Breckinridge}}, \ and\
		\bibinfo {editor} {\bibfnamefont {H.~P.}\ \bibnamefont {Stahl}}},\ \bibinfo
	{organization} {International Society for Optics and Photonics}\ (\bibinfo
	{publisher} {SPIE},\ \bibinfo {year} {2019})\ pp.\ \bibinfo {pages} {382 --
		386}\BibitemShut {NoStop}%
	\bibitem [{\citenamefont {Samantaray}\ \emph {et~al.}(2017)\citenamefont
		{Samantaray}, \citenamefont {Ruo-Berchera}, \citenamefont {Meda},\ and\
		\citenamefont {Genovese}}]{Samantaray2017}%
	\BibitemOpen
	\bibfield  {author} {\bibinfo {author} {\bibfnamefont {N.}~\bibnamefont
			{Samantaray}}, \bibinfo {author} {\bibfnamefont {I.}~\bibnamefont
			{Ruo-Berchera}}, \bibinfo {author} {\bibfnamefont {A.}~\bibnamefont {Meda}},
		\ and\ \bibinfo {author} {\bibfnamefont {M.}~\bibnamefont {Genovese}},\
	}\href {\doibase 10.1038/lsa.2017.5} {\bibfield  {journal} {\bibinfo
			{journal} {Light: Science {\&} Applications}\ }\textbf {\bibinfo {volume}
			{6}},\ \bibinfo {pages} {e17005} (\bibinfo {year} {2017})}\BibitemShut
	{NoStop}%
	\bibitem [{\citenamefont {Groom}\ \emph {et~al.}(1999)\citenamefont {Groom},
		\citenamefont {Holland}, \citenamefont {Levi}, \citenamefont {Palaio},
		\citenamefont {Perlmutter}, \citenamefont {Stover},\ and\ \citenamefont
		{Wei}}]{Groom1999}%
	\BibitemOpen
	\bibfield  {author} {\bibinfo {author} {\bibfnamefont {D.~E.}\ \bibnamefont
			{Groom}}, \bibinfo {author} {\bibfnamefont {S.~E.}\ \bibnamefont {Holland}},
		\bibinfo {author} {\bibfnamefont {M.~E.}\ \bibnamefont {Levi}}, \bibinfo
		{author} {\bibfnamefont {N.~P.}\ \bibnamefont {Palaio}}, \bibinfo {author}
		{\bibfnamefont {S.}~\bibnamefont {Perlmutter}}, \bibinfo {author}
		{\bibfnamefont {R.~J.}\ \bibnamefont {Stover}}, \ and\ \bibinfo {author}
		{\bibfnamefont {M.}~\bibnamefont {Wei}},\ }in\ \href {\doibase
		10.1117/12.347079} {\emph {\bibinfo {booktitle} {Sensors, Cameras, and
				Systems for Scientific/Industrial Applications}}},\ Vol.\ \bibinfo {volume}
	{3649},\ \bibinfo {editor} {edited by\ \bibinfo {editor} {\bibfnamefont
			{M.~M.}\ \bibnamefont {Blouke}}\ and\ \bibinfo {editor} {\bibfnamefont
			{G.~M.~W.}\ \bibnamefont {Jr.}}},\ \bibinfo {organization} {International
		Society for Optics and Photonics}\ (\bibinfo  {publisher} {SPIE},\ \bibinfo
	{year} {1999})\ pp.\ \bibinfo {pages} {80 -- 90}\BibitemShut {NoStop}%
	\bibitem [{\citenamefont {Janesick}\ and\ \citenamefont
		{of~Physics}(2007)}]{janesick2007photon}%
	\BibitemOpen
	\bibfield  {author} {\bibinfo {author} {\bibfnamefont {J.}~\bibnamefont
			{Janesick}}\ and\ \bibinfo {author} {\bibfnamefont {A.~I.}\ \bibnamefont
			{of~Physics}},\ }\href {https://books.google.com.ar/books?id=RydDzQEACAAJ}
	{\emph {\bibinfo {title} {Photon Transfer: DN [lambda]}}}\ (\bibinfo
	{publisher} {SPIE},\ \bibinfo {year} {2007})\BibitemShut {NoStop}%
	\bibitem [{\citenamefont {Brida}\ and\ \citenamefont
		{Ruo~Berchera}(2010)}]{brida_2010}%
	\BibitemOpen
	\bibfield  {author} {\bibinfo {author} {\bibfnamefont {G.}~\bibnamefont
			{Brida}}\ and\ \bibinfo {author} {\bibfnamefont {I.}~\bibnamefont
			{Ruo~Berchera}},\ }\href {\doibase 10.1038/NPHOTON.2010.29} {\bibfield
		{journal} {\bibinfo  {journal} {Nature Photonics}\ }\textbf {\bibinfo
			{volume} {4}},\ \bibinfo {pages} {227} (\bibinfo {year} {2010})}\BibitemShut
	{NoStop}%
	\bibitem [{\citenamefont {Diehl}\ \emph {et~al.}(2008)\citenamefont {Diehl},
		\citenamefont {Angstadt}, \citenamefont {Campa}, \citenamefont {Cease},
		\citenamefont {Derylo}, \citenamefont {Emes}, \citenamefont {Estrada},
		\citenamefont {Kubik}, \citenamefont {Flaugher}, \citenamefont {Holland},
		\citenamefont {Jonas}, \citenamefont {Kolbe}, \citenamefont {Krider},
		\citenamefont {Kuhlmann}, \citenamefont {Kuk}, \citenamefont {Maiorino},
		\citenamefont {Palaio}, \citenamefont {Plazas}, \citenamefont {Roe},
		\citenamefont {Scarpine}, \citenamefont {Schultz}, \citenamefont {Shaw},
		\citenamefont {Spinka},\ and\ \citenamefont {Stuermer}}]{Diehl_2008}%
	\BibitemOpen
	\bibfield  {author} {\bibinfo {author} {\bibfnamefont {H.~T.}\ \bibnamefont
			{Diehl}}, \bibinfo {author} {\bibfnamefont {R.}~\bibnamefont {Angstadt}},
		\bibinfo {author} {\bibfnamefont {J.}~\bibnamefont {Campa}}, \bibinfo
		{author} {\bibfnamefont {H.}~\bibnamefont {Cease}}, \bibinfo {author}
		{\bibfnamefont {G.}~\bibnamefont {Derylo}}, \bibinfo {author} {\bibfnamefont
			{J.~H.}\ \bibnamefont {Emes}}, \bibinfo {author} {\bibfnamefont
			{J.}~\bibnamefont {Estrada}}, \bibinfo {author} {\bibfnamefont
			{D.}~\bibnamefont {Kubik}}, \bibinfo {author} {\bibfnamefont {B.~L.}\
			\bibnamefont {Flaugher}}, \bibinfo {author} {\bibfnamefont {S.~E.}\
			\bibnamefont {Holland}}, \bibinfo {author} {\bibfnamefont {M.}~\bibnamefont
			{Jonas}}, \bibinfo {author} {\bibfnamefont {W.~F.}\ \bibnamefont {Kolbe}},
		\bibinfo {author} {\bibfnamefont {J.}~\bibnamefont {Krider}}, \bibinfo
		{author} {\bibfnamefont {S.}~\bibnamefont {Kuhlmann}}, \bibinfo {author}
		{\bibfnamefont {K.}~\bibnamefont {Kuk}}, \bibinfo {author} {\bibfnamefont
			{M.}~\bibnamefont {Maiorino}}, \bibinfo {author} {\bibfnamefont
			{N.}~\bibnamefont {Palaio}}, \bibinfo {author} {\bibfnamefont
			{A.}~\bibnamefont {Plazas}}, \bibinfo {author} {\bibfnamefont {N.~A.}\
			\bibnamefont {Roe}}, \bibinfo {author} {\bibfnamefont {V.}~\bibnamefont
			{Scarpine}}, \bibinfo {author} {\bibfnamefont {K.}~\bibnamefont {Schultz}},
		\bibinfo {author} {\bibfnamefont {T.}~\bibnamefont {Shaw}}, \bibinfo {author}
		{\bibfnamefont {H.}~\bibnamefont {Spinka}}, \ and\ \bibinfo {author}
		{\bibfnamefont {W.}~\bibnamefont {Stuermer}},\ }in\ \href {\doibase
		10.1117/12.790053} {\emph {\bibinfo {booktitle} {High Energy, Optical, and
				Infrared Detectors for Astronomy III}}},\ Vol.\ \bibinfo {volume} {7021},\
	\bibinfo {editor} {edited by\ \bibinfo {editor} {\bibfnamefont {D.~A.}\
			\bibnamefont {Dorn}}\ and\ \bibinfo {editor} {\bibfnamefont {A.~D.}\
			\bibnamefont {Holland}}},\ \bibinfo {organization} {International Society for
		Optics and Photonics}\ (\bibinfo  {publisher} {SPIE},\ \bibinfo {year}
	{2008})\ pp.\ \bibinfo {pages} {86 -- 96}\BibitemShut {NoStop}%
	\bibitem [{\citenamefont {Roundy}(1997)}]{Roundy1997Baseline}%
	\BibitemOpen
	\bibfield  {author} {\bibinfo {author} {\bibfnamefont {C.~B.}\ \bibnamefont
			{Roundy}},\ }in\ \href {\doibase 10.1117/12.281394} {\emph {\bibinfo
			{booktitle} {10th Meeting on Optical Engineering in Israel}}},\ Vol.\
	\bibinfo {volume} {3110},\ \bibinfo {editor} {edited by\ \bibinfo {editor}
		{\bibfnamefont {I.}~\bibnamefont {Shladov}}\ and\ \bibinfo {editor}
		{\bibfnamefont {S.~R.}\ \bibnamefont {Rotman}}},\ \bibinfo {organization}
	{International Society for Optics and Photonics}\ (\bibinfo  {publisher}
	{SPIE},\ \bibinfo {year} {1997})\ pp.\ \bibinfo {pages} {860 --
		879}\BibitemShut {NoStop}%
	\bibitem [{\citenamefont {Chierchie}\ \emph {et~al.}(2020)\citenamefont
		{Chierchie}, \citenamefont {Moroni}, \citenamefont {Stefanazzi},
		\citenamefont {Chavez}, \citenamefont {Paolini}, \citenamefont {Cancelo},
		\citenamefont {Haro}, \citenamefont {Tiffenberg}, \citenamefont {Estrada},\
		and\ \citenamefont {Uemura}}]{chierchie2020smartreadout}%
	\BibitemOpen
	\bibfield  {author} {\bibinfo {author} {\bibfnamefont {F.}~\bibnamefont
			{Chierchie}}, \bibinfo {author} {\bibfnamefont {G.~F.}\ \bibnamefont
			{Moroni}}, \bibinfo {author} {\bibfnamefont {L.}~\bibnamefont {Stefanazzi}},
		\bibinfo {author} {\bibfnamefont {C.}~\bibnamefont {Chavez}}, \bibinfo
		{author} {\bibfnamefont {E.}~\bibnamefont {Paolini}}, \bibinfo {author}
		{\bibfnamefont {G.}~\bibnamefont {Cancelo}}, \bibinfo {author} {\bibfnamefont
			{M.~S.}\ \bibnamefont {Haro}}, \bibinfo {author} {\bibfnamefont
			{J.}~\bibnamefont {Tiffenberg}}, \bibinfo {author} {\bibfnamefont
			{J.}~\bibnamefont {Estrada}}, \ and\ \bibinfo {author} {\bibfnamefont
			{S.}~\bibnamefont {Uemura}},\ }\href@noop {} {\enquote {\bibinfo {title}
			{Smart-readout of the skipper-ccd: Achieving sub-electron noise levels in
				regions of interest},}\ } (\bibinfo {year} {2020}),\ \Eprint
	{http://arxiv.org/abs/2012.10414} {arXiv:2012.10414 [physics.ins-det]}
	\BibitemShut {NoStop}%
	\bibitem [{Sup()}]{SupplementalMat}%
	\BibitemOpen
	\href@noop {} {\emph {\bibinfo {title} {SUPPLEMENTAL MATERIALS for: Smart
				readout of nondestructive image sensors with single-photon
				sensitivity}}}\BibitemShut {NoStop}%
	\bibitem [{\citenamefont {Cancelo}\ \emph {et~al.}(2021)\citenamefont
		{Cancelo}, \citenamefont {Chavez}, \citenamefont {Chierchie}, \citenamefont
		{Estrada}, \citenamefont {Fernandez-Moroni}, \citenamefont {Paolini},
		\citenamefont {Haro}, \citenamefont {Soto}, \citenamefont {Stefanazzi},
		\citenamefont {Tiffenberg}, \citenamefont {Treptow}, \citenamefont {Wilcer},\
		and\ \citenamefont {Zmuda}}]{cancelo2021low}%
	\BibitemOpen
	\bibfield  {author} {\bibinfo {author} {\bibfnamefont {G.~I.}\ \bibnamefont
			{Cancelo}}, \bibinfo {author} {\bibfnamefont {C.}~\bibnamefont {Chavez}},
		\bibinfo {author} {\bibfnamefont {F.}~\bibnamefont {Chierchie}}, \bibinfo
		{author} {\bibfnamefont {J.}~\bibnamefont {Estrada}}, \bibinfo {author}
		{\bibfnamefont {G.}~\bibnamefont {Fernandez-Moroni}}, \bibinfo {author}
		{\bibfnamefont {E.~E.}\ \bibnamefont {Paolini}}, \bibinfo {author}
		{\bibfnamefont {M.~S.}\ \bibnamefont {Haro}}, \bibinfo {author}
		{\bibfnamefont {A.}~\bibnamefont {Soto}}, \bibinfo {author} {\bibfnamefont
			{L.}~\bibnamefont {Stefanazzi}}, \bibinfo {author} {\bibfnamefont
			{J.}~\bibnamefont {Tiffenberg}}, \bibinfo {author} {\bibfnamefont
			{K.}~\bibnamefont {Treptow}}, \bibinfo {author} {\bibfnamefont
			{N.}~\bibnamefont {Wilcer}}, \ and\ \bibinfo {author} {\bibfnamefont {T.~J.}\
			\bibnamefont {Zmuda}},\ }\href {\doibase 10.1117/1.JATIS.7.1.015001}
	{\bibfield  {journal} {\bibinfo  {journal} {Journal of Astronomical
				Telescopes, Instruments, and Systems}\ }\textbf {\bibinfo {volume} {7}},\
		\bibinfo {pages} {1 } (\bibinfo {year} {2021})}\BibitemShut {NoStop}%
	\bibitem [{\citenamefont {Howell}(2006)}]{howell2006handbook}%
	\BibitemOpen
	\bibfield  {author} {\bibinfo {author} {\bibfnamefont {S.~B.}\ \bibnamefont
			{Howell}},\ }\href@noop {} {\emph {\bibinfo {title} {Handbook of CCD
				astronomy}}},\ Vol.~\bibinfo {volume} {5}\ (\bibinfo  {publisher} {Cambridge
		University Press},\ \bibinfo {year} {2006})\BibitemShut {NoStop}%
	\bibitem [{\citenamefont {LDSS3}(2021)}]{LDSS3}%
	\BibitemOpen
	\bibfield  {author} {\bibinfo {author} {\bibnamefont {LDSS3}},\ }\href@noop
	{} {\emph {\bibinfo {title} {The Low Dispersion Survey Spectrograph}}}
	(\bibinfo {year} {2021(accessed October 19, 2021)}),\ \bibinfo {note}
	{\url{http://www.lco.cl/wp-content/uploads/2021/02/LDSS3handout2021.pdf}}\BibitemShut
	{NoStop}%
	\bibitem [{\citenamefont {Yuan}\ \emph {et~al.}(2010)\citenamefont {Yuan},
		\citenamefont {Bao}, \citenamefont {Lu}, \citenamefont {Zhang}, \citenamefont
		{Peng},\ and\ \citenamefont {Pan}}]{YUAN20101}%
	\BibitemOpen
	\bibfield  {author} {\bibinfo {author} {\bibfnamefont {Z.-S.}\ \bibnamefont
			{Yuan}}, \bibinfo {author} {\bibfnamefont {X.-H.}\ \bibnamefont {Bao}},
		\bibinfo {author} {\bibfnamefont {C.-Y.}\ \bibnamefont {Lu}}, \bibinfo
		{author} {\bibfnamefont {J.}~\bibnamefont {Zhang}}, \bibinfo {author}
		{\bibfnamefont {C.-Z.}\ \bibnamefont {Peng}}, \ and\ \bibinfo {author}
		{\bibfnamefont {J.-W.}\ \bibnamefont {Pan}},\ }\href {\doibase
		https://doi.org/10.1016/j.physrep.2010.07.004} {\bibfield  {journal}
		{\bibinfo  {journal} {Physics Reports}\ }\textbf {\bibinfo {volume} {497}},\
		\bibinfo {pages} {1} (\bibinfo {year} {2010})}\BibitemShut {NoStop}%
	\bibitem [{\citenamefont {{Malygin}}\ \emph {et~al.}(1985)\citenamefont
		{{Malygin}}, \citenamefont {{Penin}},\ and\ \citenamefont
		{{Sergienko}}}]{malygin_1985}%
	\BibitemOpen
	\bibfield  {author} {\bibinfo {author} {\bibfnamefont {A.~A.}\ \bibnamefont
			{{Malygin}}}, \bibinfo {author} {\bibfnamefont {A.~N.}\ \bibnamefont
			{{Penin}}}, \ and\ \bibinfo {author} {\bibfnamefont {A.~V.}\ \bibnamefont
			{{Sergienko}}},\ }\href@noop {} {\bibfield  {journal} {\bibinfo  {journal}
			{Soviet Physics Doklady}\ }\textbf {\bibinfo {volume} {20}},\ \bibinfo
		{pages} {227} (\bibinfo {year} {1985})}\BibitemShut {NoStop}%
	\bibitem [{\citenamefont {Padgett}\ and\ \citenamefont
		{Boyd}(2017)}]{Padgett_2017}%
	\BibitemOpen
	\bibfield  {author} {\bibinfo {author} {\bibfnamefont {M.~J.}\ \bibnamefont
			{Padgett}}\ and\ \bibinfo {author} {\bibfnamefont {R.~W.}\ \bibnamefont
			{Boyd}},\ }\href@noop {} {\bibfield  {journal} {\bibinfo  {journal}
			{Philosophical Transactions of the Royal Society A: Mathematical, Physical
				and Engineering Sciences}\ }\textbf {\bibinfo {volume} {375}},\ \bibinfo
		{pages} {20160233} (\bibinfo {year} {2017})}\BibitemShut {NoStop}%
	\bibitem [{\citenamefont {Pittman}\ \emph {et~al.}(1995)\citenamefont
		{Pittman}, \citenamefont {Shih}, \citenamefont {Strekalov},\ and\
		\citenamefont {Sergienko}}]{Pittman_1995}%
	\BibitemOpen
	\bibfield  {author} {\bibinfo {author} {\bibfnamefont {T.~B.}\ \bibnamefont
			{Pittman}}, \bibinfo {author} {\bibfnamefont {Y.~H.}\ \bibnamefont {Shih}},
		\bibinfo {author} {\bibfnamefont {D.~V.}\ \bibnamefont {Strekalov}}, \ and\
		\bibinfo {author} {\bibfnamefont {A.~V.}\ \bibnamefont {Sergienko}},\ }\href
	{\doibase 10.1103/PhysRevA.52.R3429} {\bibfield  {journal} {\bibinfo
			{journal} {Phys. Rev. A}\ }\textbf {\bibinfo {volume} {52}},\ \bibinfo
		{pages} {R3429} (\bibinfo {year} {1995})}\BibitemShut {NoStop}%
	\bibitem [{\citenamefont {Defienne}\ \emph {et~al.}(2018)\citenamefont
		{Defienne}, \citenamefont {Reichert},\ and\ \citenamefont
		{Fleischer}}]{defienne_2010}%
	\BibitemOpen
	\bibfield  {author} {\bibinfo {author} {\bibfnamefont {H.}~\bibnamefont
			{Defienne}}, \bibinfo {author} {\bibfnamefont {M.}~\bibnamefont {Reichert}},
		\ and\ \bibinfo {author} {\bibfnamefont {J.~W.}\ \bibnamefont {Fleischer}},\
	}\href {\doibase 10.1103/PhysRevLett.120.203604} {\bibfield  {journal}
		{\bibinfo  {journal} {Phys. Rev. Lett.}\ }\textbf {\bibinfo {volume} {120}},\
		\bibinfo {pages} {203604} (\bibinfo {year} {2018})}\BibitemShut {NoStop}%
	\bibitem [{\citenamefont {Bestvater}\ \emph {et~al.}(2010)\citenamefont
		{Bestvater}, \citenamefont {Seghiri}, \citenamefont {Kang}, \citenamefont
		{Gröner}, \citenamefont {Lee}, \citenamefont {Im},\ and\ \citenamefont
		{Wachsmuth}}]{Bestvater_2010}%
	\BibitemOpen
	\bibfield  {author} {\bibinfo {author} {\bibfnamefont {F.}~\bibnamefont
			{Bestvater}}, \bibinfo {author} {\bibfnamefont {Z.}~\bibnamefont {Seghiri}},
		\bibinfo {author} {\bibfnamefont {M.}~\bibnamefont {Kang}}, \bibinfo {author}
		{\bibfnamefont {N.}~\bibnamefont {Gröner}}, \bibinfo {author} {\bibfnamefont
			{J.-Y.}\ \bibnamefont {Lee}}, \bibinfo {author} {\bibfnamefont {K.-B.}\
			\bibnamefont {Im}}, \ and\ \bibinfo {author} {\bibfnamefont {M.}~\bibnamefont
			{Wachsmuth}},\ }\href {\doibase 10.1364/OE.18.023818} {\bibfield  {journal}
		{\bibinfo  {journal} {Optics express}\ }\textbf {\bibinfo {volume} {18}},\
		\bibinfo {pages} {23818} (\bibinfo {year} {2010})}\BibitemShut {NoStop}%
	\bibitem [{\citenamefont {Orieux}\ \emph {et~al.}(2017)\citenamefont {Orieux},
		\citenamefont {Versteegh}, \citenamefont {JÃ¶ns},\ and\ \citenamefont
		{Ducci}}]{Orieux_2017}%
	\BibitemOpen
	\bibfield  {author} {\bibinfo {author} {\bibfnamefont {A.}~\bibnamefont
			{Orieux}}, \bibinfo {author} {\bibfnamefont {M.~A.~M.}\ \bibnamefont
			{Versteegh}}, \bibinfo {author} {\bibfnamefont {K.~D.}\ \bibnamefont
			{JÃ¶ns}}, \ and\ \bibinfo {author} {\bibfnamefont {S.}~\bibnamefont
			{Ducci}},\ }\href {\doibase 10.1088/1361-6633/aa6955} {\bibfield  {journal}
		{\bibinfo  {journal} {Reports on Progress in Physics}\ }\textbf {\bibinfo
			{volume} {80}},\ \bibinfo {pages} {076001} (\bibinfo {year}
		{2017})}\BibitemShut {NoStop}%
	\bibitem [{\citenamefont {Crill}\ and\ \citenamefont
		{Siegler}(2018)}]{crill2018exoplanet}%
	\BibitemOpen
	\bibfield  {author} {\bibinfo {author} {\bibfnamefont {B.}~\bibnamefont
			{Crill}}\ and\ \bibinfo {author} {\bibfnamefont {N.}~\bibnamefont
			{Siegler}},\ }\href@noop {} {\bibfield  {journal} {\bibinfo  {journal} {Jet
				Propulsion Laboratory Publications D-102506}\ } (\bibinfo {year}
		{2018})}\BibitemShut {NoStop}%
	\bibitem [{\citenamefont {Rauscher}\ \emph {et~al.}()\citenamefont {Rauscher},
		\citenamefont {Holland}, \citenamefont {Miko},\ and\ \citenamefont
		{Waczynski}}]{Rauscher_internal_2019}%
	\BibitemOpen
	\bibfield  {author} {\bibinfo {author} {\bibfnamefont {B.~J.}\ \bibnamefont
			{Rauscher}}, \bibinfo {author} {\bibfnamefont {S.~E.}\ \bibnamefont
			{Holland}}, \bibinfo {author} {\bibfnamefont {L.~R.}\ \bibnamefont {Miko}}, \
		and\ \bibinfo {author} {\bibfnamefont {A.}~\bibnamefont {Waczynski}},\
	}\href@noop {} {\ }\BibitemShut {NoStop}%
	\bibitem [{\citenamefont {{Stark}}\ \emph {et~al.}(2019)\citenamefont
		{{Stark}}, \citenamefont {{Belikov}}, \citenamefont {{Bolcar}}, \citenamefont
		{{Crill}}, \citenamefont {{Krist}}, \citenamefont {{Nemati}}, \citenamefont
		{{Pueyo}}, \citenamefont {{Rauscher}}, \citenamefont {{Riggs}}, \citenamefont
		{{Ruane}}, \citenamefont {{Sirbu}}, \citenamefont {{Soummer}}, \citenamefont
		{{St. Laurent}}, \citenamefont {{Zimmerman}},\ and\ \citenamefont
		{{Groff}}}]{Stark_2019}%
	\BibitemOpen
	\bibfield  {author} {\bibinfo {author} {\bibfnamefont {C.}~\bibnamefont
			{{Stark}}}, \bibinfo {author} {\bibfnamefont {R.}~\bibnamefont {{Belikov}}},
		\bibinfo {author} {\bibfnamefont {M.}~\bibnamefont {{Bolcar}}}, \bibinfo
		{author} {\bibfnamefont {B.}~\bibnamefont {{Crill}}}, \bibinfo {author}
		{\bibfnamefont {J.}~\bibnamefont {{Krist}}}, \bibinfo {author} {\bibfnamefont
			{B.}~\bibnamefont {{Nemati}}}, \bibinfo {author} {\bibfnamefont
			{L.}~\bibnamefont {{Pueyo}}}, \bibinfo {author} {\bibfnamefont
			{B.}~\bibnamefont {{Rauscher}}}, \bibinfo {author} {\bibfnamefont {A.~J.~E.}\
			\bibnamefont {{Riggs}}}, \bibinfo {author} {\bibfnamefont {G.}~\bibnamefont
			{{Ruane}}}, \bibinfo {author} {\bibfnamefont {D.}~\bibnamefont {{Sirbu}}},
		\bibinfo {author} {\bibfnamefont {R.}~\bibnamefont {{Soummer}}}, \bibinfo
		{author} {\bibfnamefont {K.}~\bibnamefont {{St. Laurent}}}, \bibinfo {author}
		{\bibfnamefont {N.}~\bibnamefont {{Zimmerman}}}, \ and\ \bibinfo {author}
		{\bibfnamefont {T.}~\bibnamefont {{Groff}}},\ }in\ \href@noop {} {\emph
		{\bibinfo {booktitle} {American Astronomical Society Meeting Abstracts
				\#233}}},\ \bibinfo {series} {American Astronomical Society Meeting
		Abstracts}, Vol.\ \bibinfo {volume} {233}\ (\bibinfo {year} {2019})\ p.\
	\bibinfo {pages} {402.05}\BibitemShut {NoStop}%
	\bibitem [{\citenamefont {des Poids~et Mesure}(2019)}]{BIMP2019}%
	\BibitemOpen
	\bibfield  {author} {\bibinfo {author} {\bibfnamefont {B.~I.}\ \bibnamefont
			{des Poids~et Mesure}},\ }\href@noop {} {\emph {\bibinfo {title} {Mise en
				pratique for the definition of the ampere and other electric units in the
				SI}}} (\bibinfo {year} {2019}),\ \bibinfo {note} {appendix 2}\BibitemShut
	{NoStop}%
	\bibitem [{\citenamefont {Scherer}\ and\ \citenamefont
		{Schumacher}(2019)}]{scherer2019single}%
	\BibitemOpen
	\bibfield  {author} {\bibinfo {author} {\bibfnamefont {H.}~\bibnamefont
			{Scherer}}\ and\ \bibinfo {author} {\bibfnamefont {H.~W.}\ \bibnamefont
			{Schumacher}},\ }\href@noop {} {\bibfield  {journal} {\bibinfo  {journal}
			{Annalen der Physik}\ }\textbf {\bibinfo {volume} {531}},\ \bibinfo {pages}
		{1800371} (\bibinfo {year} {2019})}\BibitemShut {NoStop}%
	\bibitem [{\citenamefont {Giblin}\ \emph {et~al.}(2012)\citenamefont {Giblin},
		\citenamefont {Kataoka}, \citenamefont {Fletcher}, \citenamefont {See},
		\citenamefont {Janssen}, \citenamefont {Griffiths}, \citenamefont {Jones},
		\citenamefont {Farrer},\ and\ \citenamefont {Ritchie}}]{giblin2012towards}%
	\BibitemOpen
	\bibfield  {author} {\bibinfo {author} {\bibfnamefont {S.}~\bibnamefont
			{Giblin}}, \bibinfo {author} {\bibfnamefont {M.}~\bibnamefont {Kataoka}},
		\bibinfo {author} {\bibfnamefont {J.}~\bibnamefont {Fletcher}}, \bibinfo
		{author} {\bibfnamefont {P.}~\bibnamefont {See}}, \bibinfo {author}
		{\bibfnamefont {T.}~\bibnamefont {Janssen}}, \bibinfo {author} {\bibfnamefont
			{J.}~\bibnamefont {Griffiths}}, \bibinfo {author} {\bibfnamefont
			{G.}~\bibnamefont {Jones}}, \bibinfo {author} {\bibfnamefont
			{I.}~\bibnamefont {Farrer}}, \ and\ \bibinfo {author} {\bibfnamefont
			{D.}~\bibnamefont {Ritchie}},\ }\href@noop {} {\bibfield  {journal} {\bibinfo
			{journal} {Nature communications}\ }\textbf {\bibinfo {volume} {3}},\
		\bibinfo {pages} {1} (\bibinfo {year} {2012})}\BibitemShut {NoStop}%
	\bibitem [{\citenamefont {Tiffenberg}\ and\ \citenamefont
		{Moroni}(2021)}]{CCDFuente}%
	\BibitemOpen
	\bibfield  {author} {\bibinfo {author} {\bibfnamefont {J.}~\bibnamefont
			{Tiffenberg}}\ and\ \bibinfo {author} {\bibfnamefont {G.~F.}\ \bibnamefont
			{Moroni}},\ }\href@noop {} {\emph {\bibinfo {title} {Pinning down the ampere
				with a supersensitive particle detector}}} (\bibinfo {year} {2020 (accessed
		April 29, 2021)}),\ \bibinfo {note}
	{{https://news.fnal.gov/2020/10/pinning-down-the-ampere-with-a-supersensitive-particle-detector/}}\BibitemShut
	{NoStop}%
\end{thebibliography}
%\bibliographystyle{apsrev4-1}


\end{document}